\documentclass[11pt]{article}		
\usepackage[super,sort,comma]{natbib}

\usepackage{booktabs}
\usepackage{array}
\usepackage{multirow}
\usepackage{caption}
\usepackage{tabularx}
\usepackage{rotating}
\usepackage{makecell}
\usepackage{ragged2e}
\usepackage{tabularx}
\usepackage{array}
\usepackage{booktabs}
\usepackage{array, tabularx, booktabs, ragged2e}
\usepackage[table]{xcolor}   

\usepackage{amsmath} 
\usepackage{fancyhdr}		

\usepackage[section]{placeins}   %

\usepackage{graphicx}

\makeatletter \renewcommand\@biblabel[1]{$^{#1}$} \makeatother
\usepackage{xcolor}

\newcommand{\filename}[1]{%
  \colorbox{gray!15}{\texttt{\detokenize{#1}}}%
}

\newcommand{\cen}[1]{\begin{center} #1 \end{center}}

\usepackage[mathlines]{lineno}

\usepackage{hyperref}
\hypersetup{ colorlinks,
    citecolor=blue,
    filecolor=blue,
    linkcolor=blue,
    urlcolor=blue
}

\usepackage{xcolor}

\definecolor{gray}{rgb}{0.6,0.6,0.6}
\definecolor{red}{rgb}{0.85,0,0}
\definecolor{green}{rgb}{0,0.85,0}
\definecolor{blue}{rgb}{0,0,0.85}
\definecolor{beige}{rgb}{0.92,0.87,0.78}
\usepackage[all]{hypcap}    

\begin{document}

\newcommand{\affilmark}[1]{\textsuperscript{#1}}

\cen{\sf
{\Large \bfseries
COBRA2026: a large-scale multicenter pelvic cone-beam computed tomography projection dataset
}
\vspace*{1mm}

{\small
Adrian Thummerer\affilmark{1,2,3,4},
Simon Rit\affilmark{5},
Florian Kamp\affilmark{6},
Matteo Maspero\affilmark{7,8},
Martjin P.W. Intven\affilmark{7},
Thomas G. Boné\affilmark{9},
Christopher Kurz\affilmark{4,10,11},
Guillaume Landry\affilmark{4,10,11},
Thomas Baudier\affilmark{5},
Mustafa Kadhim\affilmark{12,13},
Julius Arnold\affilmark{14,15},
Michael Rauter\affilmark{16},
Barbara Knäusl\affilmark{14,15},
Lukas Zimmermann\affilmark{14,15}
}
\vspace{2mm}

{\scriptsize
\vspace{0.2em}
\affilmark{1}Department of Radiation Oncology, Inselspital, Bern University Hospital and University of Bern, Bern, Switzerland \\
\affilmark{2}Department of Digital Medicine, Faculty of Medicine, University of Bern, Bern, Switzerland \\
\affilmark{3}Sitem Center for Translational Medicine and Biomedical Entrepreneurship, University of Bern, Bern, Switzerland\\
\affilmark{4}Department of Radiation Oncology, LMU University Hospital, LMU Medizin, LMU Munich\\
\affilmark{5}INSA‐Lyon, Université Lyon 1, CNRS, Inserm, CREATIS UMR 5220, U1294, F‐69373, Lyon, France\\
\affilmark{6}Department of Radiation Oncology and Cyberknife Center, University Hospital of Cologne, Cologne, Germany\\
\affilmark{7}Department of Radiotherapy, University Medical Center Utrecht, Utrecht, Netherlands \\
\affilmark{8}Computational Imaging Group for MR Diagnostics \& Therapy, University Medical Center Utrecht, Utrecht, The Netherlands\\
\affilmark{9}Department of Therapeutic Radiology and Oncology, Medical University of Graz, Graz, Austria\\
\affilmark{10}German Cancer Consortium (DKTK), partner site Munich, a partnership between DKFZ and LMU University Hospital Munich, Germany\\
\affilmark{11}Bavarian Cancer Research Center (BZKF), Munich, Germany \\
\affilmark{12}Department of Medical Radiation Physics, Lund University, Lund, Sweden \\
\affilmark{13}Radiation Physics, Department of Hematology, Oncology, and Radiation Physics, Skåne University Hospital, Lund, Sweden \\
\affilmark{14}Department of Radiation Oncology, Medical University of Vienna, Vienna, Austria \\
\affilmark{15}Christian Doppler Laboratory for Image and Knowledge Driven Precision Radiation
Oncology, Medical University of Vienna, Vienna, Austria \\
\affilmark{16}University of Applied Sciences Wiener Neustadt, Competence Center for Preclinical
Imaging and Biomedical Engineering, Wiener Neustadt, Austria\\
}

\vspace{2mm}
{\small Version typeset \today\\
Corresponding author: Adrian Thummerer (adrian.thummerer@unibe.ch)} \\
}

\pagenumbering{roman}
\setcounter{page}{1}
\pagestyle{plain}

\begin{abstract}

\noindent {\bf Purpose:} The COBRA2026 dataset is a large-scale multicenter dataset of raw radiotherapy cone-beam computed tomography (CBCT) acquisitions designed for the development and benchmarking of CBCT reconstruction and image-correction methods. The dataset targets researchers working on image reconstruction, medical image processing, and deep learning-based methods for image-guided and adaptive radiotherapy.\

\noindent{\bf Acquisition and Validation Methods:} Raw CBCT projection data, acquisition geometry, calibration and correction information, clinically reconstructed CBCT images, and corresponding planning CT images were collected retrospectively for 867 patients undergoing pelvic radiotherapy at six European centers. Acquisitions were obtained using Elekta and Varian imaging systems. Vendor-specific files were anonymized and converted into open formats. Planning CT images were deformably registered to the daily CBCT anatomy, and matched projections were simulated using the corresponding acquisition geometry. All cases were visually reviewed for completeness and registration quality, and cases with substantial processing or registration errors (large displacements of organs) were excluded.\

\noindent{\bf Data Format and Usage Notes:} The approximately 950-GB dataset comprises the finally selected 867 cases divided into training, validation, and test sets containing 692, 52, and 123 cases, respectively. Projection stacks and volumetric images are provided as compressed MetaImage files, while acquisition geometry and metadata are provided in XML and YAML formats. The dataset is released under the CC BY-NC 4.0 license, indexed on Zenodo (\href{https://doi.org/10.5281/zenodo.21322350}{doi.org/10.5281/zenodo.21322350}), and hosted in an institutional data repository. Pre-processing and baseline reconstruction scripts are openly available in git repositories.

\noindent{\bf Potential Applications:} The COBRA2026 dataset supports research on conventional and deep learning-based CBCT reconstruction, sparse-view and low-dose imaging, artifact and scatter correction, motion compensation, and synthetic CT generation. Furthermore, the COBRA2026 dataset is the basis for the COBRA2026 challenge (\href{https://cobra2026.grand-challenge.org/}{cobra2026.grand-challenge.org}).

\end{abstract}



\newpage

\setlength{\baselineskip}{0.7cm}      

\pagenumbering{arabic}
\setcounter{page}{1}
\pagestyle{fancy}
\section{Introduction}
High-precision radiotherapy relies on accurate imaging throughout the treatment workflow, from diagnosis and target delineation to treatment planning, daily patient positioning, and treatment verification and adaptation in response to anatomical changes \citep{dawson2007advances, jaffray2002flat}. Cone-beam computed tomography (CBCT) is widely used for daily image guidance \citep{landry2018current, posiewnik2019review}. CBCT systems acquire a sequence of two-dimensional X-ray projections while rotating around the patient and reconstruct these projections into a three-dimensional volumetric image. Owing to their compact design, CBCT systems can be integrated directly into treatment devices, including linear accelerators and proton therapy gantries, thereby providing volumetric images of the patient in treatment position immediately before irradiation \citep{herrick2023systematic}. However, compared with planning CT, CBCT image quality is degraded by effects such as X-ray scatter, beam hardening, detector lag, truncation, and patient motion \citep{nagarajappa2015artifacts}. These effects reduce Hounsfield unit accuracy and limit the use of CBCT for quantitative applications, including automated organ segmentation and dose calculation \citep{giacometti2020review, spadea2021deep, liang2023segmentation}.

A broad range of approaches has been investigated to reduce the image-quality gap between CBCT and planning CT. These include hardware developments, such as anti-scatter grids and improved detector technology \citep{stankovic2017optimal, lustermans2024image}; advanced analytical and iterative reconstruction methods \citep{mory2016motion, wang2009iterative}; and, more recently, deep learning-based techniques \cite{altalib2025synthetic, thummerer2020comparison}. Although learning-based methods have shown considerable promise, most studies have concentrated on the image domain by converting reconstructed CBCTs into synthetic images resembling CTs\citep{spadea2021deep}. Comparatively less attention has been given to projection-domain methods, which offer the potential to account more directly for the underlying acquisition physics and sources of image degradation \citep{nomura2019projection}.

Progress in both image- and projection-domain research is constrained by the limited availability of suitable data. Deep learning methods, in particular, benefit from large and diverse datasets. However, publicly available radiotherapy CBCT datasets remain scarce, and recently released large scale datasets, including SynthRAD2023 and SynthRAD2025, provide reconstructed images but not the raw projection and acquisition data required to develop or validate reconstruction algorithms \citep{hugo2017longitudinal, thummerer2023synthrad2023, thummerer2025synthrad2025}. A further challenge is the lack of geometrically consistent reference data. Real CBCT acquisitions cannot be paired with perfectly matched planning CT data because patient anatomy and positioning typically differ between the planning and the image-guided treatment time points. Consequently, many studies rely entirely on simulated projections, for which the extent of the domain gap relative to clinical measurements remain uncertain \citep{shieh2019spare}.

To address these limitations, we present COBRA2026 (Cone-Beam Reconstruction for Radiotherapy Applications), a large-scale, multicenter dataset containing real CBCT projection data acquired at a range of linear accelerators using different CBCT systems, planning CT images deformed to the anatomy of the corresponding CBCT and simulated CBCT projections generated from the deformed CT. 
This combination provides both the realism of clinical measurements and the geometrically aligned reference data required for supervised learning and quantitative evaluation.

The COBRA2026 dataset is intended to support research across a broad range of tasks, including sparse-view and low-dose CBCT reconstruction, scatter correction, artifact reduction, and synthetic CT generation. The dataset is publicly available under the Creative Commons Attribution-NonCommercial 4.0 International (CC BY-NC 4.0) license and will form the basis of the COBRA2026 deep learning challenge, which aims to stimulate the development and systematic evaluation of methods for high-quality CBCT reconstruction.

\section{Acquisition and Validation Methods}

\subsection{Dataset overview}
\begin{figure}[htbp]
    \centering
    \includegraphics[width=1\textwidth]{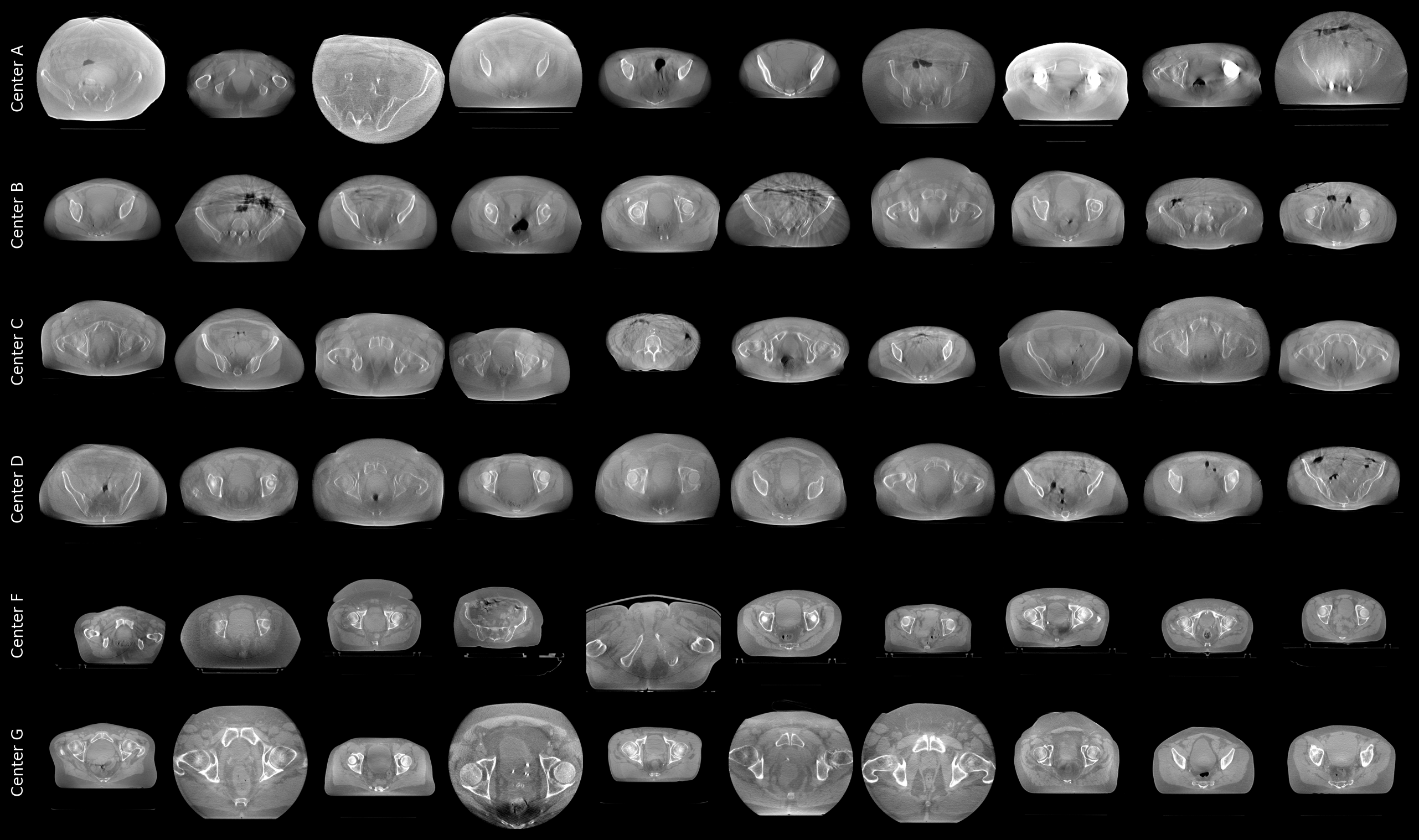}
    \footnotesize
    \caption{Representative axial pelvic CBCT images from the six participating centers, illustrating variability in patient anatomy, field of view, image quality, acquisition characteristics, and imaging artifacts. CBCTs from centers A-D are displayed with a window/level of 800/-75 HU and from centers F and G with 1200/250 HU.}
    \label{fig:overview}
\end{figure}

The COBRA2026 dataset comprises raw CBCT acquisition data, including projections, acquisition geometry, and calibration data; corresponding planning CTs; and simulated CBCT projections generated from the planning CTs of 867 patients undergoing radiation therapy for pelvic malignancies. The data were collected at six European radiotherapy centers: Department of Radiation Oncology, University Medical Center Utrecht, The Netherlands; Department of Radiation Oncology, Medical University of Vienna, Austria; Department of Radiation Oncology, Centre Léon Bérard, Lyon, France; Department of Radiation Oncology, LMU University Hospital Munich, Germany; Department of Radiation Oncology, University Hospital Cologne, Germany; and Department of Therapeutic Radiology and Oncology, Medical University Graz, Austria. \autoref{fig:overview} presents axial slices of reconstructed CBCTs from 10 cases per center to visualize the variability of the dataset.

The CBCTs were acquired using imaging systems integrated into linear accelerators from two manufacturers: Elekta AB (Stockholm, Sweden) and Varian Medical Systems, Inc.--A Siemens Healthineers Company (Palo Alto, California, USA). Detailed overviews of the CBCT and CT acquisition parameters are provided in Tables~\ref{tab:cbct_params} and~\ref{tab:ct_params}, respectively. Ethical approval for the retrospective collection and sharing of the data was obtained from the responsible ethics committee at each participating center (UMC Utrecht: 22/474, Medical University Vienna: 1904/2025, Centre Léon Bérard Lyon: R201-000-030, LMU Munich: 24-0195, Medical University Graz: 1620/2025, and University Hospital Cologne: 24-120\_1-retro).

\begin{table}[htbp]
\footnotesize
\centering
\renewcommand{\arraystretch}{1.3}
\caption{CBCT acquisition and reconstruction parameters by center. Per-patient metadata is available in the dataset.}
\label{tab:cbct_params}
\begin{tabular}{@{}>{\raggedright\arraybackslash}p{2.4cm}
    *{6}{>{\raggedright\arraybackslash}p{1.9cm}}@{}}
\toprule
\textbf{Parameter} & \textbf{Center A} & \textbf{Center B} & \textbf{Center C} & \textbf{Center D} & \textbf{Center F} & \textbf{Center G} \\
\midrule
\textbf{Manufacturer} & Elekta & Elekta & Elekta & Elekta & Varian & Varian \\
\textbf{Linac} & Synergy, VersaHD & Synergy, VersaHD & Synergy & Synergy, VersaHD & Halcyon, TrueBeam & TrueBeam, VitalBeam \\
\textbf{Tube voltage [$kV$]} & 120 & 120 & 120 & 120 & 125--140 & 125--140 \\
\textbf{Tube current [$mA$]} & 20--80 & 40 & 40 & 40 & 30--90 & 30--76 \\
\textbf{Pulse length [$ms$]} & 32--40 & 40 & 40 & 40 & 14.3--20 & 20--25 \\
\textbf{Arc} & Full (142),\newline Half (8) & Full & Full & Full & Full (98),\newline Half (18) & Full (111),\newline Half (56) \\
\textbf{Scatter grid} & No & No & No & No & Yes & Yes \\
\textbf{Projection resolution [$mm^2$]} & 0.8 $\times$ 0.8 & 0.8 $\times$ 0.8 & 0.8 $\times$ 0.8 & 0.8 $\times$ 0.8 & 0.28--0.39 $\times$ 0.39--1.12 & 0.28--0.39 $\times$ 0.39--1.12 \\
\textbf{Projection size} & 504 $\times$ 504 & 504 $\times$ 504 & 504 $\times$ 504 & 504 $\times$ 504 & 1024--3072 $\times$ 384--768 & 1024--1536 $\times$ 384--768 \\
\textbf{Projection number} & 189--499 & 630--734 & 338--730 & 346--693 & 406--895 & 498--895 \\
\textbf{Offset detector} & Yes (129),\newline No (21) & Yes & Yes & Yes & Yes (18),\newline No (98) & Yes (111),\newline No (56) \\
\textbf{Reconstruction spacing [$mm^2$]} & 1 $\times$ 1 & 1 $\times$ 1 & 1 $\times$ 1 & 1 $\times$ 1 & 0.55--1.05 $\times$ 0.55--1.05 & 0.51--0.95 $\times$ 0.51--0.95 \\
\textbf{Slice thickness [$mm$]} & 1 & 1 & 1 & 1 & 2 & 2 \\
\bottomrule
\end{tabular}
\end{table}

\begin{table}[htbp]
\vspace{1em}
\footnotesize
\centering
\renewcommand{\arraystretch}{1.3}
\caption{CT acquisition and reconstruction parameters by center. Per-patient metadata is available in the dataset.}
\label{tab:ct_params}
\begin{tabular}{@{}>{\raggedright\arraybackslash}p{2.4cm}
    *{6}{>{\raggedright\arraybackslash}p{1.9cm}}@{}}
\toprule
\textbf{Parameter} & \textbf{Center A} & \textbf{Center B} & \textbf{Center C} & \textbf{Center D} & \textbf{Center F} & \textbf{Center G} \\
\midrule
\textbf{Manufacturer} & Philips,\newline Siemens & Siemens & Philips,\newline Siemens & TOSHIBA & Siemens & Canon Medical Systems,\newline Philips,\newline TOSHIBA \\
\textbf{Model} & Big Bore,\newline Biograph40 & Somatom Definition AS & Brilliance Big Bore,\newline GEMINI TF Big Bore,\newline SOMATOM Confidence,\newline SOMATOM Definition Edge & Aquillion/LB & SOMATOM go.Sim & Aquillion Exceed LB,\newline Big Bore,\newline Aquillion/LB \\
\textbf{Tube voltage [$kV$]} & 120 & 120 & 120--140 & 120 & 110--120 & 120 \\
\textbf{Tube current [$mA$]} & 42--332 & 36--276 & 40--321 & 40--228 & 13--328 & 88--484 \\
\textbf{Pulse length [$ms$]} & 469--1238 & 500 & 500--3195 & 500--750 & 438--1250 & 500--1000 \\
\textbf{CTDI$_{\text{vol}}$} & 1.9--20.7 & 2.47--18.99 & 8.2--42.5 & 5.3--42.9 & 0.98--31.5 & 2.8--29.3 \\
\textbf{Rows/Columns} & 512/512 & 512/512 & 512/512 & 512/512 & 512/512 & 512/512 \\
\textbf{Pixel spacing [$mm^2$]} & 0.74--1.37 $\times$ 0.74--1.37 & 0.87--1.27 $\times$ 0.87--1.27 & 0.86--1.37 $\times$ 0.86--1.37 & 1.07--1.37 $\times$ 1.07--1.37 & 1.17 $\times$ 1.17 & 1.07--1.37 $\times$ 1.07--1.37 \\
\textbf{Slice thickness [$mm$]} & 2--3 & 2--4 & 1--3 & 1--3 & 1--3 & 1--3 \\
\textbf{Data collection diameter [$mm$]} & 500--700 & 500 & 500--600 & 550--700 & 600.5 & 550--700 \\
\textbf{Reconstruction diameter [$mm$]} & 379--700 & 447--650 & 441--700 & 550--700 & 600 & 550--700 \\
\textbf{Age} & 24--91 & 32--92 (64.4) & 29--90 (66.4) & N/A & N/A & N/A \\
\textbf{Sex} & M (77),\newline F (73) & F (75),\newline M (59) & N/A & N/A & M (88),\newline F (28) & N/A \\
\bottomrule
\end{tabular}
\vspace{1em}
\end{table}

\subsection{Pre-processing pipeline}
Pre-processing was performed to anonymize and harmonize the data collected from the participating centers, convert it into openly accessible, vendor-neutral file formats, construct matched CT--CBCT pairs, and generate simulated projections from the corresponding CTs. To preserve the characteristics of the clinically acquired data, the raw CBCT acquisitions were only converted to a vendor-neutral format, and all available files associated with each acquisition were retained and included in the release. No further image processing was performed on the raw acquisition data.

The complete pre-processing pipeline relies exclusively on open-source software and is publicly available in a GitHub repository \citep{thummerer_preprocess}. The pipeline can also be used to process additional datasets in a manner consistent with the data released in this study. The main pre-processing steps are summarized below, while detailed instructions for executing the pipeline are provided in the repository.

\subsubsection{Data conversion}
\label{sec}

In the first stage of pre-processing, the CBCT projections, acquisition geometry, calibration data, clinically reconstructed CBCT, and planning CT (pCT) were converted from vendor-specific or proprietary formats into open, vendor-neutral formats commonly used in medical image processing.

For Elekta acquisitions (Centers A--D), each center exported the raw projection files (\filename{.his}), acquisition geometry (\filename{Frames.xml}), and accompanying acquisition and reconstruction metadata (\filename{.INI}). The individual \filename{.his} files were stacked into a single three-dimensional \filename{projections.mha} file without conversion to line integral using the Elekta projection reader provided by the Reconstruction Toolkit (RTK, v2.7.0)~\citep{rit2024reconstruction}. The resulting projection stack contains the measured X-ray detector signal stored as 16-bit unsigned integers with a nominal range of 0--65,535. The acquisition geometry was extracted from \filename{Frames.xml}, converted into an RTK geometry object using the RTK Elekta geometry reader, and stored as \filename{geometry.xml}. Acquisition- and reconstruction-related parameters were extracted from the accompanying \filename{.INI} files and stored in \filename{reconstruction.yaml}. The available parameters differed among centers and all fields containing patient-identifying information were removed before release.

For Varian acquisitions (Centers F and G), each center exported the raw projection images (\filename{.xim}), a file containing scan metadata and geometry information (\filename{Scan.xml}), and a \filename{Calibrations} directory containing acquisition-specific calibration and correction data. The individual \filename{.xim} files were stacked into a single \filename{projections.mha} file wihtout conversion to line integrals using the RTK Varian projection reader. The acquisition geometry was reconstructed from the \filename{Scan.xml} file and the headers of the \filename{*.xim} files using the RTK Varian geometry reader and stored as \filename{geometry.xml}.

In addition to the CBCT acquisition data, the clinically reconstructed CBCT and the corresponding pCTs were exported from the clinical systems and converted from DICOM to \filename{.mha} using the SimpleITK python package (v3.0.0a1). General acquisition and imaging metadata were extracted from the pCT DICOM headers, projection headers, and vendor-specific files, \filename{.INI} for Elekta and \filename{Scan.xml} for Varian, and consolidated into a single \filename{metadata.yaml} file for each patient.

\subsubsection{CBCT reconstruction}

To provide a consistently reconstructed reference volume for image registration, the projection data were reconstructed independently of the vendor software using the open-source RTK implementation of the Feldkamp--Davis--Kress (FDK) algorithm~\cite{rit2014reconstruction,feldkamp1984practical}. Before reconstruction, the raw detector signal $I$ was normalized using a flood-field estimate $I_0$ and converted into line-integral measurements $p$ according to

\begin{equation}
p = -\ln\left(\frac{I}{I_0}\right)
\label{eq}
\end{equation}

For Elekta acquisitions, the exported projection images had already undergone flood-field correction prior to export. Consequently, no additional pixel-wise flood-field correction was applied. The scalar normalization value \filename{FloodImageFilterNorm} provided in the corresponding \filename{*.INI} files was instead used to normalize the projection intensities and was scaled according to the exposure of the respective acquisition using \autoref{eq:flood_field_scaling}:

\begin{equation}
I_0
=
\mathrm{FloodImageFilterNorm} \cdot
\frac{\mathrm{mA}_{\mathrm{scan}} \cdot \mathrm{ms}_{\mathrm{scan}}}
     {\mathrm{mA}_{\mathrm{air}} \cdot \mathrm{ms}_{\mathrm{air}}}.
\label{eq:flood_field_scaling}
\end{equation}
where \(\mathrm{mA}_{\mathrm{scan}}\) and \(\mathrm{mA}_{\mathrm{air}}\) denote the tube currents, and \(\mathrm{ms}_{\mathrm{scan}}\) and \(\mathrm{ms}_{\mathrm{air}}\) denote the exposure times, used for the patient and air-scan acquisitions, respectively.

For Varian acquisitions, air-scan projections acquired at a limited number of gantry angles, typically approximately ten per scan, were available in the \filename{Calibrations} directory. Each patient's projection was normalized using the air scan acquired at the closest gantry angle, after scaling the air-scan signal to account for the acquisition-specific tube current--time product.
Furthermore, for Varian acquisitions only, a kernel-based scatter-correction method, following the approach described by Sun et al.\citep{sun2010improved}, was additionally applied before converting the normalized detector signal into line integrals. This additional correction was required because reconstructions generated without scatter correction exhibited pronounced low-frequency shading artifacts near the center of the image. An equivalent correction was not applied to the Elekta acquisitions because the required scatter-model parameters were unavailable and comparable artifacts were not observed in the reconstructed images.

The RTK FDK reconstruction incorporated displaced-detector weighting to account for offset-detector, or half-fan, acquisition geometry and Parker short-scan weighting to account for the incomplete gantry rotation \citep{parker1982optimal, wang2002x}. A ramp filter using a Hann window was applied during filtered backprojection (cut frequency for Elekta acquisitions 0.99, for Varian acquisitions 0.4)\citep{harris1978windows}.

Each CBCT volume was reconstructed using the voxel spacing and matrix dimensions of the corresponding clinically reconstructed CBCT. The reconstructed attenuation values ($\mathrm{CBCT_{\mu}}$) were subsequently converted to CT-Numbers ($\mathrm{CBCT_{HU}}$, approximate Hounsfield units) according to the conversion described by Park et al.\citep{park2015proton}:

\begin{equation} 
\mathrm{CBCT_{HU}} = \mathrm{CBCT_{\mu}} \cdot 2^{16} - 1024 
\label{eq:hu_conversion} 
\end{equation}

A CBCT field-of-view (FOV) mask was generated for each acquisition. For Elekta acquisitions, the mask was derived by thresholding the reconstructed image, whereas for Varian acquisitions, it was calculated from the acquisition geometry using the RTK FOV Filter.

\subsubsection{Deformable image registration}
\label{sec:dir}
To obtain a pCT that was geometrically consistent with the daily anatomy depicted in the CBCT, the pCT was registered to the independently reconstructed CBCT using the Elastix framework \citep{klein2009elastix, marstal2016simpleelastix}. The pCT was first rigidly aligned with the CBCT and was subsequently deformably registered using a B-spline transformation optimized with the IMPACT similarity metric \citep{boussot2025impact} yielding a deformed CT. Rather than directly comparing CT and CBCT intensities, IMPACT compares feature representations extracted using a pretrained neural network (TotalSegmentator layer M730), thereby improving robustness to residual modality-dependent intensity differences and CBCT artifacts. Registration was performed using a multi-resolution strategy. The complete registration parameter files are provided in the public code repository.

Because deformable image registration alone could not reliably resolve differences in the location and extent of intestinal gas pockets, an additional air-cavity matching procedure was applied. For this correction, residual CBCT intensity artifacts were reduced by generating a deep learning-based synthetic CT using a model trained on simulated data of a single center; a link to the corresponding model parameters is provided in \autoref{tab:resources}. Air-cavity masks were obtained by intensity thresholding using thresholds of $-400$~HU for Elekta acquisitions and $-200$~HU for Varian acquisitions. Regions containing mismatched air cavities were corrected either by replacing erroneous air pockets in the deformed CT with soft-tissue intensities or by transferring the corresponding air cavities from the synthetic CT to the deformed CT. To avoid sharp edges between modified regions, boundaries were blended using Gaussian smoothing.

The planning CTs from Center B were acquired after administration of a contrast agent. To better approximate the non-contrast appearance of the treatment-time CBCT, contrast enhancement in the deformed CT was suppressed within the available bowel-bag contour.

The original pCT generally extended beyond the CBCT reconstruction FOV. To retain the complete patient anatomy required for forward projection, regions outside the CBCT FOV were incorporated into the deformed CT using an extended deformation vector field. A smooth transition was applied between the registered region within the CBCT FOV and the surrounding pCT anatomy to avoid discontinuities and sharp edges in the resulting volume.

\subsubsection{Projection simulation}

To provide projection data paired with a known and geometrically consistent reference anatomy, cone-beam projections were simulated from the deformed CT using the acquisition geometry of the corresponding clinical scan. An acquisition was simulated for every gantry angle recorded in \filename{geometry.xml} using the simcbctgenerator package (see \autoref{tab:resources})\citep{zimmermann2026eliminating}. Thereby matching the angular sampling and scanner geometry of the measured acquisition.

To approximate intra-scan motion encountered in clinical pelvic CBCT acquisitions, a randomized respiratory motion model for the pelvis was applied during forward projection. Patient-specific motion amplitudes between 2 and 7~mm and motion frequencies between 12 and 20~cycles per minute were sampled randomly. The resulting time-dependent transformations were applied during projection generation displacing the abdominal wall so that individual projections represented different phases of the simulated motion trajectory. This motion simulation replicates realistic streaking artifacts in air--tissue interfaces.

Bowtie filters are defined for both Elekta and Varian machines using air scans, resulting in a spatially varying primary photon count $N_{\mathrm{primary}}$, derived from the physics parameters and approximated using SpekPy \citep{poludniowski2021spekpy}. The spectral photon fluence at the detector plane was then generated using a tungsten anode with a 14$^\circ$ target angle, evaluated at a source-to-detector distance of the respective scan. Vendor-specific beam filtration was applied: for Elekta, a 13.5 mm aluminium flat filter; for Varian, a 2.5 mm aluminium inherent tube filtration followed by a 0.4 mm titanium flat filter. The incident count per pixel was obtained as $N_0 = bp \cdot mAs \cdot \phi \cdot A_{pixel} \cdot g$, where $\phi$ is the SpekPy fluence, $A_{pixel}$ the pixel area, $bp$ the modelled beam profile (bowtie/flat-field response received by flat-field measurement fits), and $g$ a vendor-specific gain factor (0.851 for Elekta, 2.65 for Varian) that reconciles the computed fluence with empirically calibrated reference exposures and accounts for differences in detector quantum efficiency and scintillator response. The primary count in each pixel was finally given by $N_{primary} = N_0 \cdot e^{-p}$, where $p$ is the line integral of the attenuation from the forward projection, after which scatter was added and Poisson noise was applied to model quantum statistics.

\subsubsection{Couch removal and spatial origin}

In the final pre-processing stage, the patient-support couch was automatically segmented in the reconstructed CBCT and removed from the CBCT FOV mask, resulting in a couch-free patient mask. Image geometry was represented in an LPS patient coordinate system, in which the positive (x)-, (y)-, and (z)-axes point towards the patient’s left, posterior, and superior directions, respectively. The spatial origins of all released volumetric images were subsequently shifted such that they align with the CBCT acquisition isocenter. This transformation established a consistent spatial reference frame across patients, acquisition systems, and participating centers. The origin transformation was not applied to the original CT or the clinically reconstructed CBCT. The center of the clinical reconstruction was not always aligned with the center of the projections, hence its spatial extent may differ from that of the deformed CT and the corresponding FOV mask.

\begin{figure}[htbp]
    \centering
    \includegraphics[width=1\textwidth]{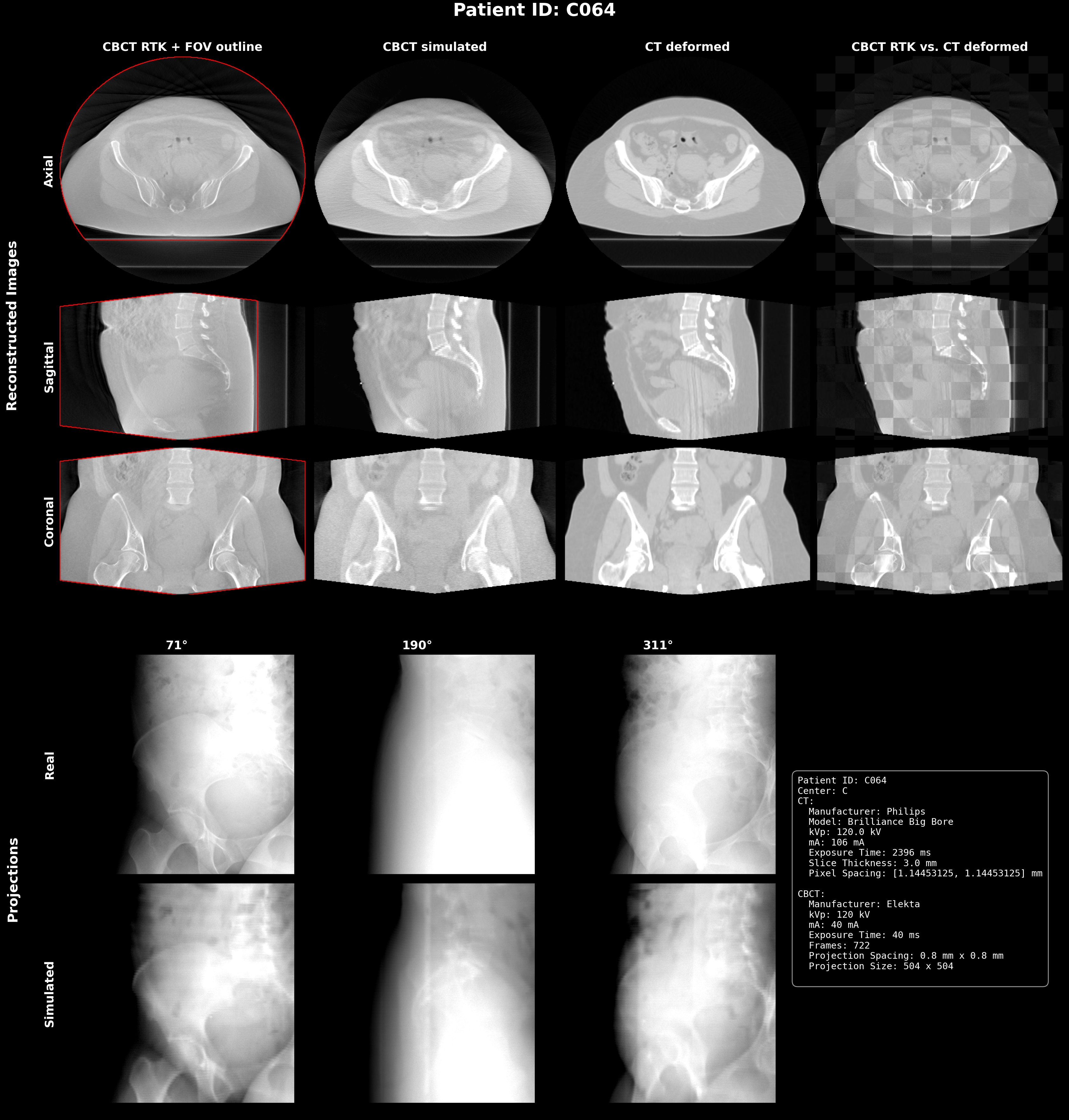}
    \caption{Overview of case C064 showing axial (first row), sagittal (second row), and coronal (third row) views. Columns (left to right): RTK- CBCT reconstructed from real projections, RTK-CBCT reconstructed from simulated projections, and deformed planning CT. Additionally, the last two rows allow comparison of the simulated and acquired projections at three different angles. Reconstructed images are presented with a window/ level of 1368/-340 HU, projections (log transformed) with a dynamic window based on the 5th and 95th intensity percentile (per projection).}
    \label{fig:example}
\end{figure}

For visual quality control, a multi-panel overview figure was generated for each patient. The figure includes the clinically reconstructed CBCT, the independently reconstructed CBCT, the deformed CT, and side-by-side comparisons of representative measured and simulated projections. An example is presented in Figure~\ref{fig:example}

\subsubsection{Generated files}

Following completion of the pre-processing pipeline, the files listed in Table~\ref{tab:generated_files} were generated for each patient and included in the public COBRA2026 dataset. 

\begin{table}[htbp]
\vspace{1em}
\centering
\footnotesize
\caption{Files generated by the pre-processing pipeline and included in the COBRA dataset for each patient.}
\label{tab:generated_files}
\begin{tabular}{p{0.32\linewidth} p{0.60\linewidth}}
\hline
\textbf{File} & \textbf{Description} \\
\hline
\multicolumn{2}{l}{\textit{Raw CBCT data and geometry}} \\
\filename{projections.mha} & Stacked raw CBCT projection images \\
\filename{geometry.xml} & Cone-beam acquisition geometry in RTK format \\
\filename{projections_simulated.mha} & Cone-beam projections simulated from the deformed CT \\
\hline
\multicolumn{2}{l}{\textit{Image volumes}} \\
\filename{ct_original.mha} & Planning CT, without any registration or corrections. \\
\filename{ct_def.mha} & Deformed planning CT, full FOV. \\
\filename{ct_def_masked.mha} & Deformed planning CT, masked to CBCT FOV. \\
\filename{cbct_clinical.mha} & Vendor-reconstructed clinical CBCT. \\
\filename{cbct_rtk.mha} & CBCT reconstructed using RTK FDK. \\
\hline
\multicolumn{2}{l}{\textit{Masks}} \\
\filename{fov_cbct.mha} & CBCT FOV mask. \\
\filename{fov_cbct_nocouch.mha} & CBCT FOV mask without couch. \\
\hline
\multicolumn{2}{l}{\textit{Metadata and vendor files}} \\
\filename{metadata.yaml} & Acquisition parameters extracted from CT, projections, and headers. \\
\filename{reconstruction.yaml} & Acquisition and reconstruction metadata from \filename{*.INI} files (Elekta only). \\
\filename{Scan.xml} & Acquisition, geometry and reconstruction metadata (Varian only). \\
\filename{Calibrations/} & Calibration and Correction parameter files (Varian only). \\
\hline
\multicolumn{2}{l}{\textit{Quality control}} \\
\filename{overview_<ID>.png} & Multi-panel overview figure displaying main files. \\
\hline
\end{tabular}
\vspace{1em}
\end{table}

\subsection{Data validation}

The COBRA2026 dataset was designed to support research on CBCT reconstruction and image correction using a heterogeneous sample of patients treated at multiple European radiotherapy centers. The inclusion criteria were intentionally broad, with eligibility primarily determined by the availability of pelvic CBCT projection data and a corresponding planning CT. The dataset was not filtered or balanced according to age, sex, or cancer indication. Cases containing clinically encountered artifacts or devices, including metal implants and rectal balloons, were intentionally retained to reflect the variability of routine clinical imaging. Initially, around 200 patients per center were collected.

Following pre-processing, all cases were visually reviewed for data completeness and deformable registration quality using the automatically generated overview figures (\autoref{fig:example}). Registration quality was assessed using checkerboard overlays of the independently reconstructed CBCT and deformed CT in the central axial, coronal, and sagittal planes. In addition, measured and simulated projections were compared at three representative gantry angles.

Cases with substantial deformable registration errors or incomplete data were excluded from the initial selection. Consequently, the number of retained cases differed among participating centers. The remaining cases were divided into training, validation, and test sets with consideration of registration quality: cases assigned to the validation and test sets were required to show high anatomical agreement between the reconstructed CBCT and the deformed CT to ensure that quantitative evaluation was based on geometrically reliable reference data.

\section{Data Format and Usage Notes}

\subsection{Dataset structure and file formats}

To facilitate organization of the COBRA2026 challenge, the dataset was divided into training, validation, and test sets. Table~\ref{tab:data_split} summarizes the number of cases contributed by each center to the respective subsets. A unique identifier was assigned to each case consisting of a center identifier followed by a three-digit case number, for example, \filename{D152}. The top-level dataset directory contains separate subdirectories for the training, validation, and test sets. Within each subset, cases are grouped by center, with a separate subdirectory for each case. Each case directory contains the files listed in Table~\ref{tab:generated_files}.

Volumetric images and projection stacks are provided as compressed MetaImage files (\filename{.mha}), which can be read, written, and visualized using commonly available open-source medical image-processing software based on the Insight Toolkit (ITK)\citep{mccormick2014itk}. Acquisition geometry and metadata are provided in \filename{.xml} and \filename{.yaml} formats, respectively.

\begin{table}[htbp]
\footnotesize
\centering
\renewcommand{\arraystretch}{1.3}
\caption{Dataset split for training, validation and testing sets}
\label{tab:data_split}
\begin{tabular}{@{}>{\raggedright\arraybackslash}p{1.7cm}|
    *{6}{>{\centering\arraybackslash}p{1.7cm}}
    |>{\centering\arraybackslash}p{1.7cm}@{}}
\toprule
\textbf{} & \textbf{Center A} & \textbf{Center B} & \textbf{Center C} & \textbf{Center D} & \textbf{Center F} & \textbf{Center G} & \textbf{Total} \\
\midrule
\textbf{Training}   & 120   & 107   & 120   & 120   & 92    & 133   & 692\\
\textbf{Validation} & 9     & 8     & 9     & 9     & 7     & 10    & 52\\
\textbf{Testing}    & 21    & 19    & 21    & 21    & 17    & 24    & 123\\
\midrule
\textbf{Total}      & 150   & 134   & 150   & 150   & 116   & 167   & 867\\
\bottomrule
\end{tabular}
\vspace{2em} 
\end{table}

\subsection{License and release}

The COBRA2026 dataset is released under the Creative Commons Attribution--NonCommercial 4.0 International (CC BY-NC 4.0) license. This license permits users to share and adapt the data for non-commercial purposes, provided that appropriate attribution is given. The complete license terms are available at \url{https://creativecommons.org/licenses/by-nc/4.0/}.

Due to the large size of the dataset, approximately 950~GB, the data are hosted in a repository operated by CREATIS (Lyon). A persistent digital object identifier (DOI) and a corresponding dataset record are provided through Zenodo.

The training set was released in July 2026. The validation and test sets will remain under embargo until March 2032 to support the fair and independent evaluation of submissions to the COBRA2026 deep learning challenge. The dataset download locations and release dates for the individual subsets are summarized in Table~\ref{tab:resources}.

\begin{table}[htbp]
\vspace{1em}
\footnotesize
\centering
\renewcommand{\arraystretch}{1.3}
\caption{Overview of COBRA2026 dataset releases and associated resources.}
\label{tab:resources}

\begin{tabular}{@{}
    >{\raggedright\arraybackslash}p{3.3cm}
    >{\centering\arraybackslash}p{2.8cm}
    >{\centering\arraybackslash}p{4.0cm}
    >{\centering\arraybackslash}p{2.3cm}
    >{\centering\arraybackslash}p{2.1cm}
    @{}}
\toprule
\textbf{Resource}
& \textbf{Link}
& \textbf{DOI}
& \textbf{Release date}
& \textbf{Comments} \\
\midrule
\textbf{Training set}
    & \href{https://tomoradio-warehouse.creatis.insa-lyon.fr/#collection/696f75f74c0b0d3d4bc7019d/folder/69b93ca75363131190cefb54}{Download Link}
    & \href{https://doi.org/10.5281/zenodo.21322350}{10.5281/zenodo.21322350}
    & July 2026
    & Size: 748 GB \\
\textbf{Validation set}
    & \href{https://tomoradio-warehouse.creatis.insa-lyon.fr/#collection/696f75f74c0b0d3d4bc7019d/folder/69b93cd55363131190cefb55}{Download Link}
    & \href{https://doi.org/10.5281/zenodo.21322350}{10.5281/zenodo.21322350}
    & March 2032
    & Size: 60 GB \\
\textbf{Testing set}
    & \href{https://tomoradio-warehouse.creatis.insa-lyon.fr/#collection/696f75f74c0b0d3d4bc7019d/folder/69b93ce25363131190cefb56}{Download Link}
    & \href{https://doi.org/10.5281/zenodo.21322350}{10.5281/zenodo.21322350}
    & March 2032
    & Size: 140 GB \\
\midrule
\textbf{Preprocessing}
    & \href{https://github.com/cobra-challenge-2026/preprocessing}{GitHub repository}
    & \href{https://doi.org/10.5281/zenodo.21397774}{10.5281/zenodo.21397774}
    & July 2026
    & - \\
\textbf{Challenge tools} 
    & \href{https://github.com/cobra-challenge-2026/gc-tools}{GitHub repository}
    & \href{https://doi.org/10.5281/zenodo.21397869}{10.5281/zenodo.21397869}
    & July 2026
    & - \\
\textbf{simcbctgenerator}
    & \href{https://github.com/openvoxelmed/simcbctgenerator}{GitHub repository}
    & \href{https://doi.org/10.48550/arXiv.2602.02130}{10.48550/arXiv.2602.02130}
    & February 2026
    & - \\
\textbf{sCT model params.}
    & \href{https://huggingface.co/zimmeryWo/simcbctgenerator-sct-model-PELVIS}{Model parameters}
    & -
    & July 2026
    & used in \ref{sec:dir} \\
\bottomrule
\end{tabular}

\vspace{2em}
\end{table}

\subsection{Challenge data conversion scripts}

The COBRA2026 dataset underpins the COBRA2026  challenge, which focuses on CBCT reconstruction. This challenge will be hosted on the Grand Challenge platform and organized in collaboration with the MIDL 2027 conference. To facilitate participation and meet the platform's technical requirements, supplementary data conversion procedures are necessary. Scripts for converting the released dataset into a format compatible with the challenge are made available in a separate public code repository (see Table~\ref{tab:resources}) \citep{thummerer_gctools}.

On the challenge platform projections are provided already converted into line integrals (using Eq.~\ref{eq}) and with raw detector signals normalized using the corresponding flood-field estimate, $I_0$, according to the vendor-specific procedures described in Section~\ref{sec}. For Varian acquisitions, the before mentioned kernel-based scatter correction is additionally applied. Finally, the acquisition geometry and metadata are converted from \filename{.xml} and \filename{.yaml}, respectively, into \filename{.json} files compatible with the Grand Challenge platform. On the platform participants will have access only to the subset of input files listed in Table~\ref{tab:gc_files}.

\begin{table}[htbp]
\vspace{1em}
\centering
\footnotesize
\caption{Files generated by the grand-challenge conversion script.}
\label{tab:gc_files}
\begin{tabular}{p{0.32\linewidth} p{0.60\linewidth}}
\hline
\textbf{File} & \textbf{Description} \\
\hline
\multicolumn{2}{l}{\textit{Inputs (available for participants on grand-challenge.org)}} \\
\filename{cbct_projections.mha} & Real CBCT projections (line integrals, corrections applied) \\
\filename{cbct_projections_simulated.mha} & Simulated CBCT projections (line integrals) \\
\filename{cbct_geometry.json} & CBCT geometry saved as \filename{.json} \\
\filename{cbct_fov.mha} & CBCT FOV mask, excluding the patient couch \\
\filename{cbct_metadata.json} & CBCT metadata file saved as \filename{.json}  \\
\hline
\multicolumn{2}{l}{\textit{Ground Truth (used for evaluation on grand-challenge.org)}} \\
\filename{ct.mha} & Deformed planning CT, masked to CBCT FOV \\
\filename{cbct.mha} & Reconstructed CBCT (measured projections). \\ 
\hline
\end{tabular}
\end{table}

\subsection{Reconstruction scripts}

Baseline scripts to perform a simple FDK reconstruction are provided in the challenge tools GitHub repository \citep{thummerer_gctools}. This repository provides the necessary utilities to read the geometry.json file, apply flood field and scatter correction (Varian only), to reconstruct the CBCT on the same grid and with the same origin as the reference CT or FOV mask, and to convert linear attenuation coefficients to CT numbers (see \autoref{eq:hu_conversion}).

\section{Discussion}

The COBRA2026 dataset provides a large-scale resource for developing and evaluating CBCT reconstruction and image-correction methods in radiotherapy, which has not previously been readily available at this scale \citep{shieh2019spare,thummerer2023synthrad2023,thummerer2025synthrad2025}. It combines raw clinical CBCT projection data with acquisition geometry, calibration and correction information, clinically reconstructed CBCT images, deformably aligned planning CT images, and simulated projections. The inclusion of data from multiple European radiotherapy centers and different imaging systems captures variety in scanner geometry, detector characteristics, acquisition protocols, and clinical image quality. Together with the COBRA2026 challenge, the dataset is intended to facilitate the development, comparison, and reproducible benchmarking of conventional and learning-based reconstruction approaches.

\subsection{Comparison with existing public datasets}

Table~\ref{tab:public_cbct_datasets} compares COBRA2026 with other publicly accessible CBCT datasets. Existing radiotherapy datasets primarily support longitudinal imaging, image registration, segmentation, or synthetic CT generation and, with the exception of SPARE, do not include the underlying projection measurements \citep{hugo2017longitudinal,thummerer2023synthrad2023,thummerer2025synthrad2025,hong2022ct}. These datasets are therefore valuable for image-domain research but cannot directly support the development or evaluation of projection-domain reconstruction and correction methods. SPARE provides measured CBCT projection data and acquisition geometry but contains only 11 measured clinical acquisitions, supplemented by 12 simulated cases, and focuses specifically on sparse-view thoracic 4D-CBCT reconstruction \citep{shieh2019spare}. COBRA2026 substantially extends the amount and diversity of clinical projection data available for radiotherapy research. 

Outside radiotherapy, additional public CBCT datasets provide complementary resources for the development of reconstruction and image-analysis methods. The Walnuts dataset includes measured laboratory cone-beam projections acquired using multiple acquisition geometries \citep{der2019cone}, whereas the ICASSP 3D-CBCT challenge provides simulated low-dose and clinical-dose projection data at a substantially larger scale \citep{biguri2024advancing}. ToothFairy2 provides a large collection of annotated dental CBCT volumes but does not include projection measurements \citep{bolelli2026multi}.

\newcolumntype{L}[1]{>{\RaggedRight\arraybackslash}p{#1}}
\newcolumntype{Y}[1]{>{\hsize=#1\hsize\RaggedRight\arraybackslash}X}

\newcolumntype{Y}[1]{%
  >{\hsize=#1\hsize\RaggedRight\arraybackslash\hyphenpenalty=10000\relax}X}
\newcolumntype{L}[1]{%
  >{\RaggedRight\arraybackslash\hyphenpenalty=10000\relax}p{#1}}
\newcommand{\access}[3][]{%
  \ifx\relax#1\relax #2\else\href{#1}{#2}\fi\newline #3}

\begin{table}[htbp]
\centering
\caption{Comparison of publicly accessible radiotherapy CBCT datasets and selected related CBCT datasets. Scale is reported using the unit given by the original publication and is therefore not directly comparable across datasets. Proj.\ data: \emph{None} indicates that only reconstructed images are released; \emph{sim.}~$=$~simulated, \emph{meas.}~$=$~measured.}
\label{tab:public_cbct_datasets}
\scriptsize
\setlength{\tabcolsep}{3.5pt}
\renewcommand{\arraystretch}{1.1}
\begin{tabularx}{\textwidth}{@{}
  L{2.6cm}
  Y{1.05}
  Y{0.85}
  Y{0.55}
  Y{1.55}
  Y{1.00}
@{}}
\toprule
\textbf{Dataset} & \textbf{Domain/ Application} & \textbf{Scale} &
\textbf{Proj.\ data} & \textbf{Additional data} & \textbf{Access/License} \\
\midrule
\multicolumn{6}{@{}l}{{\textbf{A. Radiotherapy CBCT datasets}}} \\
\addlinespace[2pt]
4D-Lung (2017) \citep{hugo2017longitudinal} &
Thorax; longitudinal 4D imaging and registration &
20 patients; 507 4D-CBCT scans &
none &
Longitudinal 4DCT and 4D-CBCT images; target and organ contours &
\access[https://www.cancerimagingarchive.net/collection/4d-lung/]{TCIA}{CC BY 3.0}\\
\addlinespace[2pt]
SPARE (2019) \citep{shieh2019spare} &
Thorax; sparse-view 4D-CBCT reconstruction &
23 cases (12 sim., 11 meas.) &
meas. and sim. &
Planning 4DCT; acquisition geometry; respiratory-phase information &
\access[https://ses.library.usyd.edu.au/handle/2123/34378]{Institutional}{CC BY-NC 4.0}\\
\addlinespace[2pt]
SynthRAD2023 \citep{thummerer2023synthrad2023} &
Brain, pelvis; synthetic CT generation &
540 CBCT--CT pairs; multicenter &
none &
Reconstructed CBCT; planning CT; masks &
\access[https://zenodo.org/records/7260705]{Zenodo}{CC BY-NC 4.0} \\
\addlinespace[2pt]
SynthRAD2025 \citep{thummerer2025synthrad2025} &
Head/neck, thorax, abdomen; synthetic CT generation &
1472 CBCT--CT pairs; multicenter &
none &
Reconstructed CBCT and planning CT pairs &
\access[https://zenodo.org/records/14918089]{Zenodo}{CC BY-NC 4.0}\\
\addlinespace[2pt]
Pancreatic CBCT (2022) \citep{hong2022ct} &
Upper abdomen; registration, organ segmentation &
40 patients; 2 CBCTs each &
none &
Planning CT; CBCT; expert OAR contours &
\access[https://www.cancerimagingarchive.net/collection/pancreatic-ct-cbct-seg/]{TCIA}{CC BY 4.0} \\
\addlinespace[3pt]
\rowcolor{black!7}
\textbf{COBRA2026} (this work) &
Pelvis; reconstruction, correction, synthetic CT generation &
867 patients; multicenter &
meas. and sim. &
Acquisition geometry and metadata; calibration/correction data;
original and deformed planning CT&
\access[https://zenodo.org/records/21322350]{Zenodo/Institutional}{CC BY-NC 4.0} \\
\midrule
\multicolumn{6}{@{}l}{\textbf{B. Other cone-beam CT datasets (selection)}} \\
\addlinespace[2pt]
Walnuts (2019) \citep{der2019cone} &
Laboratory CBCT; high-cone-angle reconstruction; deep learning research &
42 objects; 3 orbits each &
meas. &
Full acquisition geometry; reference reconstructions; reconstruction code &
\access[https://zenodo.org/records/2686726]{Zenodo}{CC BY 4.0}\\
\addlinespace[2pt]
ToothFairy2 (2024) \citep{bolelli2026multi} &
Dental/maxillofacial; multi-structure segmentation &
480 volumes &
none &
Expert annotations of 42 dental and maxillofacial structures &
\access[https://ditto.ing.unimore.it/toothfairy2/]{Institutional}{CC BY-SA}\\
\addlinespace[2pt]
ICASSP 3D-CBCT challenge (2024) \citep{biguri2024advancing} &
Simulated chest; low-dose reconstruction &
1010 sim.\ volumes &
sim. &
Two simulated dose levels; reference CT volumes &
\access[https://zenodo.org/records/8379858]{Zenodo}{CC BY 4.0} \\
\bottomrule
\end{tabularx}
\end{table}

\subsection{Strengths}

A major strength of the COBRA2026 dataset lies in its extensive scale and heterogeneity. The dataset encompasses 867 clinical cases collected at six radiotherapy centers utilizing systems from two major manufacturers. It embodies considerable variation in detector panels, acquisition geometries, reconstruction fields of view, and imaging protocols. This diversity enables evaluation of whether reconstruction algorithms generalize effectively across institutions, scanner configurations, and acquisition settings, rather than performing well only on data from a single system or protocol.

An additional advantage is the comprehensive detail and breadth of information provided for each case. The release comprises raw projection data, acquisition geometry, available calibration and correction files, the planning CT, and the clinically reconstructed CBCT. The original acquisition data were only modified when necessary for anonymization and conversion into vendor-neutral file formats. Consequently, the released files accurately mirror the data exported from the clinical imaging and treatment systems.

Another distinctive aspect is the availability of both measured and simulated projections. The simulated projections offer geometrically consistent image--projection pairs suitable for supervised learning and quantitative evaluation. Meanwhile, the measured projections retain the noise, scatter, detector response, motion artifacts, and system-specific anomalies characteristic of clinical acquisitions. Both sets of projections utilize the corresponding recorded acquisition geometry, thereby minimizing discrepancies related to angular sampling, detector configuration, and scanner geometry.

Finally, the data and associated processing tools are provided in open, vendor-neutral formats. The pipelines for conversion, reconstruction, registration, and challenge-preparation are based on public open-source software. This approach endorses reproducibility, enables users to inspect and modify individual processing steps, and facilitates the conversion of additional institutional datasets into a structure compatible with COBRA2026.

\subsection{Limitations}

Several limitations of the dataset should be considered. First, the geometrically aligned CT reference images were generated using deformable image registration and therefore do not represent exact ground-truth representations of the CBCT anatomy. Registration was particularly challenging in the presence of substantial anatomical changes, low CBCT contrast, truncation, metal artifacts, rectal balloons, differences in intestinal gas distribution, and detector saturation. Although the registrations were visually reviewed and cases with major registration errors were excluded, residual local misalignments remain. These were particularly apparent at the body contour, which was difficult to reproduce when substantial differences existed between the CBCT and planning CT, for example because of changes in skin folds or the position of the external male genitalia.

The additional air-cavity matching procedure improved the anatomical consistency between the deformed CT and CBCT but could also introduce localized artifacts when low-intensity regions in the CBCT were incorrectly classified as air. Such misclassifications occurred in a small number of cases, particularly in regions affected by streak artifacts or adjacent to high-density bone structures. The corrected air cavities should therefore be regarded as approximations of the treatment-time gas distribution. In addition, we provide the original CT alongside the air-cavity matched and deformed CT.

The simulated projections represent a simplified approximation of the clinical acquisition process. Although the recorded acquisition geometry and a randomized motion model were used, the simulation does not reproduce all physical effects present in real CBCT acquisitions \citep{zimmermann2026eliminating}. Furthermore, the effective spatial resolution of the simulated projections is limited by the resolution of the planning CT images used for forward projection, some of which were reconstructed with slice thicknesses of up to 4~mm. Consequently, the simulated projections contain less high-frequency anatomical detail than the measured detector data.

For Varian acquisitions, an implementation of a kernel-based scatter-correction method is provided and used to generate the files required for the challenge platform. The method was introduced to reduce visible low-frequency shading artifacts in the reconstructed Varian data and uses Varian specific calibration information following the methodology described by Sun et al. \cite{sun2010improved}. However, because the original implementation was not publicly available, the provided implementation represents an independent approximation and may differ from the originally described method in several respects. Residual scatter-related artifacts may therefore remain in the corrected projections and reconstructed images.

The selection of validation and test cases, based in part on deformable registration quality, introduces a potential selection bias. Requiring close anatomical agreement between the CBCT and deformed CT provides more accurate reference data for quantitative evaluation, but may favor cases with less pronounced anatomical changes, fewer artifacts, or otherwise more favorable image quality. The validation and test sets may therefore be less representative of the complete clinical cohort than the training set.

Finally, the dataset is also restricted to pelvic imaging. Although the pelvic region exhibits substantial anatomical variability and several challenging sources of CBCT artifacts, results obtained using the COBRA2026 dataset may not directly generalize to other anatomical sites, such as the head and neck, thorax, or upper abdomen. These regions differ in their motion patterns, attenuation characteristics, fields of view, and artifact distributions. Future extensions could therefore include additional anatomical regions, imaging systems, and acquisition protocols.

\subsection{Implications for radiotherapy}

CBCT is a central imaging modality in image-guided radiotherapy and is routinely used for patient position verification before treatment delivery \citep{jaffray2002flat,dawson2007advances}. Improvements in CBCT reconstruction quality may therefore have a direct effect on several components of the radiotherapy workflow, but especially on adaptive radiotherapy, in which treatment plans are modified in response to changes in patient anatomy \citep{lim2017online,sonke2019adaptive}. Reliable adaptation decisions depend on accurate representation of the patient anatomy at the time of treatment, but currently is still often hindered by impaired image quality and quantitative accuracy of conventional CBCT \citep{giacometti2020review,rusanov2022deep}. Additional image correction, synthetic CT generation, or acquisition of a separate diagnostic-quality CT before dose recalculation or treatment adaptation is therefore often required \citep{giacometti2020review,o2022assessment,thummerer2020comparison}. Improved correction methods that operate directly in the projection domain may be able to address some of the underlying causes of image degradation before these artifacts are propagated into the reconstructed volume \citep{rusanov2022deep,nomura2019projection,kurz2016investigating}.

The availability of raw projection data also enables investigation of tasks that cannot be adequately studied using reconstructed images alone. Potential applications include sparse-view reconstruction, projection subsampling \citep{chan2024minimum}, low-dose imaging \citep{thummerer2025deep} , scatter correction \citep{nomura2019projection}, truncation correction \citep{cai2021reducing}, metal-artifact reduction \citep{washio2020metal}, motion-compensated reconstruction \citep{rit2009fly}, and reconstruction from incomplete or corrupted measurements.

\section{Conclusion}

The COBRA2026 dataset provides a large-scale, multicenter collection of raw clinical CBCT projection data, acquisition geometry, calibration information, reconstructed CBCT images, planning CT images and geometrically matched simulated projections. By combining clinically realistic measurements with corresponding CT-based reference data, the dataset supports the development and systematic evaluation of projection-domain reconstruction and image-correction methods. The use of open, vendor-neutral file formats and publicly available processing tools further promotes reproducibility and facilitates extension of the dataset with additional institutional data. Together with the COBRA2026 deep learning challenge, this resource is intended to accelerate research toward more accurate and quantitatively reliable CBCT imaging for image-guided and adaptive radiotherapy.

\section{Acknowledgments}
The LMU dataset collection was supported by the BZKF Lighthouse Image Guidance in Local Therapies and the Deutsche Krebshilfe project “Deep learning for dose reduction and auto-segmentation in CBCT-guided online adaptive radiotherapy” (project number 70114849). The financial support by the Austrian Federal Ministry of Economy, Energy and Tourism, the National Foundation for Research, Technology and Development and the Christian Doppler Research Association is gratefully acknowledged.
\clearpage


\section*{References}
\small
\addcontentsline{toc}{section}{\numberline{}References}
\vspace*{-12mm}





\bibliography{./medphys_dois}      



\bibliographystyle{./medphy-doi.bst}    


\end{document}